\documentclass[11pt,twoside]{article}

\usepackage{asp2006}
\usepackage{epsf}
\usepackage{psfig}
\usepackage{lscape}

\begin{document}
\title{Physical and Wind Properties of [WC] Stars}   
%%% Fill in title
\author{Paul A. Crowther}   
%%% Fill in author names
\affil{Dept of Physics \& Astronomy, University of Sheffield, Hounsfield
Rd, Sheffield, S3 7RH, United Kingdom}    
%%% Fill in author affiliations

\begin{abstract}
We review the properties of carbon-sequence ([WC]) Wolf-Rayet 
central stars of planetary nebulae (CSPNe). 
Differences  between the subtype distribution of [WC] stars and their 
massive WC cousins are discussed. We conclude that [WO]-type differ
from early-type [WC] stars as a result of weaker stellar winds due to high 
surface gravities, and that late- and early-type [WC] and
[WO] stars generally span a  similar range in abundances, 
X(He)$\sim$X(C)$\gg$X(O), consistent with a late thermal pulse, and 
likely progenitors to PG1159 stars.
\end{abstract}

%%% MAIN BODY OF TEXT GOES HERE. CONSULT "INSTRUCTIONS FOR AUTHORS USING
%%% LATEX2E MARKUP", SECTIONS 2.3-2.6 FOR HELP WITH EQUATIONS, FIGURES,
%%% AND TABLES.

\section{Introduction}   %%% Top level section head (remove "%" symbol)

This review discusses the properties of the small fraction of central 
stars of Planetary Nebulae (CSPNe) which share a spectroscopic appearance 
with massive, carbon-sequence (WC-type) Wolf-Rayet stars. Massive WC 
stars are the chemically evolved descendents of  initially very massive O 
stars  ($M_{\rm init} \geq 25 M_{\odot}$) exhibiting the C and O products 
of 
core helium burning, plus a unique emission line spectral 
appearance due to fast, dense stellar wind outflows. Such stars are
young, with ages of only a few Myr of which several hundred cases
are known within the Milky Way, supplemented by thousands more known
in external star-forming galaxies (Crowther 2007; Hamann these proc.). 

CSPNe possessing
a similar spectral morphology to WC stars are denoted [WC] and are at
a post-Asymptotic Giant Branch (AGB) phase in the late stages of
evolution of low or intermediate mass stars ($M_{\rm init} \sim 1-5 
M_{\odot}$?) with only a few dozen examples known in the Milky Way plus
a handful in the Magellanic Clouds. With respect to normal H-rich
CSPNe, the unusual surface chemical composition of [WC] stars apparently
results from a late thermal pulse (LTP), causing a H-deficient surface, 
and likely connection with other H-deficient stars, most notably PG1159 
stars (Werner et al. these proc.)

\section{Spectral classification}

\begin{figure}[ht!] 
\plotfiddle{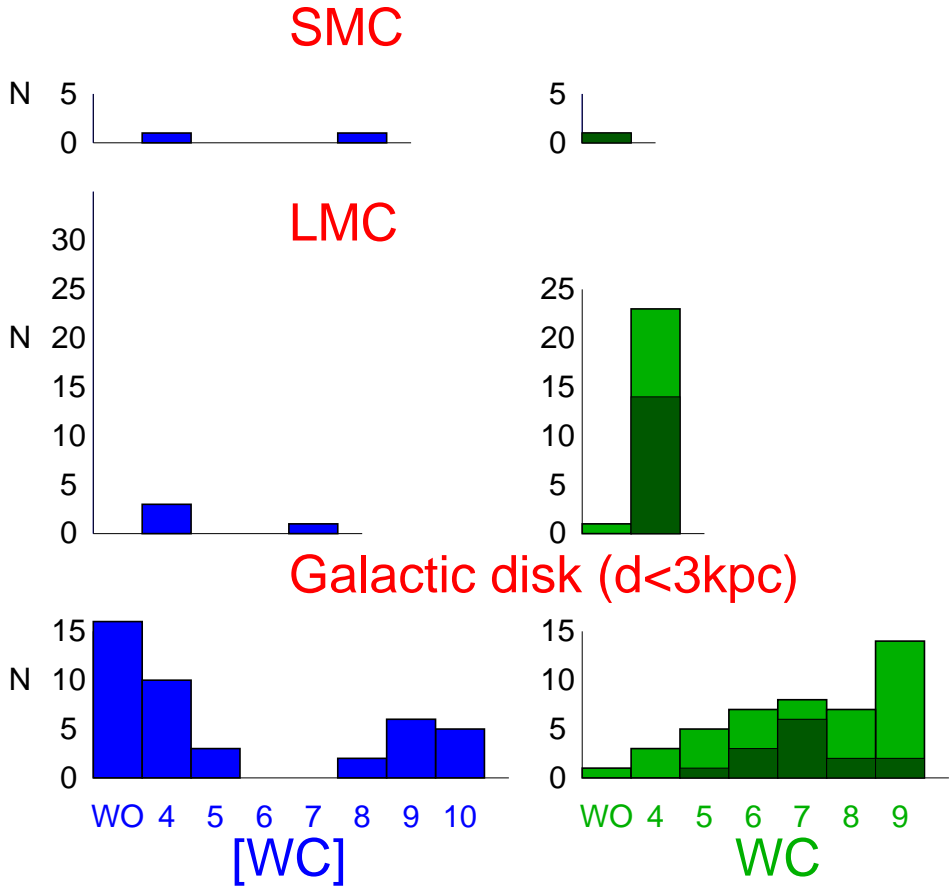}{11cm}{0}{65}{65}{-180}{-10} 
\caption{Differences in spectral subtype distribution between massive
WC and WO stars in the Solar neighbourhood ($\leq$ 3kpc) 
plus Magellanic Clouds from Crowther (2007), and WR-type CSPN in the 
Galactic disk, drawn from the CSPNe catalogue
except for updates from Crowther et al. (1998), Acker \& Neiner (2003).
We have excluded weak emission line stars and hybrid [WO]-PG1159
stars from these statistics}
\label{wc_pop} 
\end{figure}

Visual spectral classification of WR stars is based on emission
line strengths follows the nomenclature introduced by
C.S. Beals and H.H. Plaskett for nitrogen-rich (WN) and
carbon-rich (WC) stars, later updated by Smith (1968). An oxygen-rich
(WO) subclass was introduced by Barlow \& Hummer (1982). The most recent 
unified scheme spans WC4 to WC11, and WO1 to WO4, based upon the relative 
strengths of C\,{\sc iii-iv} and O\,{\sc v-vi} and C\,{\sc iv}, 
respectively (Crowther, De Marco, 
\& Barlow 1998).  It is well known that the subtype distribution of 
massive WC and WO stars differs from CSPNe, in the sense that the Milky 
Way is dominated by intermediate WC7$\pm$2 subtypes, while
CSPNe peak at high and low ionization, i.e. [WO] and WC9--10. In 
addition, WC subtype distributions depend upon their host
galaxy metallicity, with WC4 subtypes dominant in the metal-poor LMC.
These differences are illustrated in Fig.~\ref{wc_pop}, drawn from
Crowther (2007) for massive WC stars, and the CSPNe
catalogue\footnote{
http://www.arm.ac.uk/\~\,csj/research/catalogue/wccatalog.html}, 
except for updates from Crowther et al. (1998), Acker \& 
Neiner (2003). The latter classification scheme is specifically applied 
to just CSPNe, so caution has to be used when inter-comparing [WC] and
and WC subtype distributions. Galactic bulge [WC]-type CSPNe have
been omitted due to difficulties with obtaining robust spectral 
classifications, in the sense that high dust extinctions hinder
observations in the blue for the diagnostic O~{\sc vi} 3811-34 doublet.

% 1 x wc11 (K2-16)
% 5 x wc10 (CPD, IRAS07027, M4-18, He2-113, Vo1*)
% 6 x wc9 (he2-459, he2-99, bd+30, he2-142*, pe1-7*) + swst1 (wc9pec)
% 2 x wc8 (ngc 40, m2-43)
% 3 x wc5 (m1-25, m2-20*, m3-15*)
% 10 x wc4 (ngc6751, ngc5315, m1-60, pm 1-89*, H1-29*, M1-32*
%          m1-51*, Cn1-5*, m2-31*, ngc 6369*)
% 16 x wo (ngc6905,ngc7027, ic1747, ngc 1501, pb6, ngc5189, sand 3
%          ngc 2867*, he2-55*, ngc 2452*, Hb4*, PC14*,
%          m3-30*, he2-429*, ic 1297*, he2-436*)
%
% omitted He-61*, V605*, m2-4, ic4776, ngc 2371, abell 30/78, ngc6578

Regardless of classification issues, it is apparent that some WC-type
stars are morphologically identical to [WC] stars, e.g. the comparison 
between Campbell's star BD+30$^{\circ}$ 3639 and HD~164270 by Smith \& 
Aller (1971).
In general, [WC] stars possess lower wind velocities, although this
is not a defining characteristic. Indeed, the only defining characteristic
amongst Milky Way stars is the presence of a bright nebula for [WC] stars.

The [WO] sequence stars overlap with PG1159 stars (Smith \& Aller 1969)
in the sense that the latter often shows weak O\,{\sc vi} 3811-34 
emission.
Intermediate [WO]-PG1159 stars include NGC~2371 and Abell 78. Beyond [WC], 
[WO] and PG1159 stars, a number of emission line CSPNe have been grouped 
together as `weak emission line stars' (wels), although this represents a
heterogeneous group, often dictated to by the properties of the nebula.
For example, a very strong nebular continuum might cause a genuine [WO] 
star to be mis-labelled as a weak emission line CSPNe. Overall, care
should be taken to distinguish between the various flavours of emission
line CSPNe, since some more closely resemble Of stars (e.g. IRAS 
21282+5050, Crowther et al. 1998) than [WC] stars.

\section{Physical properties}

Our interpretation of hot, luminous stars via radiative transfer codes is
hindered with respect to normal stars by several effects. First, the
routine assumption of LTE breaks down for high-temperature stars. In
non-LTE, the determination of populations uses rates which are functions
of the radiation field, itself a function of the populations.
Consequently, it is necessary to solve for the radiation field and
populations iteratively.
Second, the problem of accounting for the effect of millions of
spectral lines upon the emergent atmospheric structure and emergent
spectrum -- known as line blanketing -- remains challenging for stars in
which spherical, rather than plane-parallel, geometry must be assumed due 
to
stellar winds, since the scale height of their atmospheres is not
negligible with respect to their stellar radii. The combination of
non-LTE, line blanketing (and availability of atomic data thereof), and 
spherical geometry has prevented the routine analysis of such stars until 
recently.

Specifically for Wolf-Rayet stars, radiative transfer is generally solved 
in  the co-moving frame, using either the CMFGEN (Hillier \& Miller 1998) 
and PoWR (Gr\"{a}fener, Koesterke, \& Hamann 2002) codes. In general,
the majority of published [WC] studies have employed earlier 
non-blanketed versions of these codes, applied to late-type [WC] stars
(Leuenhagen, Hamann, \& Jeffrey 1996; Leuenhagen \& Hamann 1998; De Marco 
\&
 Crowther 1998) and early-type [WC] and [WO] stars (Koesterke
\& Hamann 1997ab).

\begin{figure}[ht!]
\plotfiddle{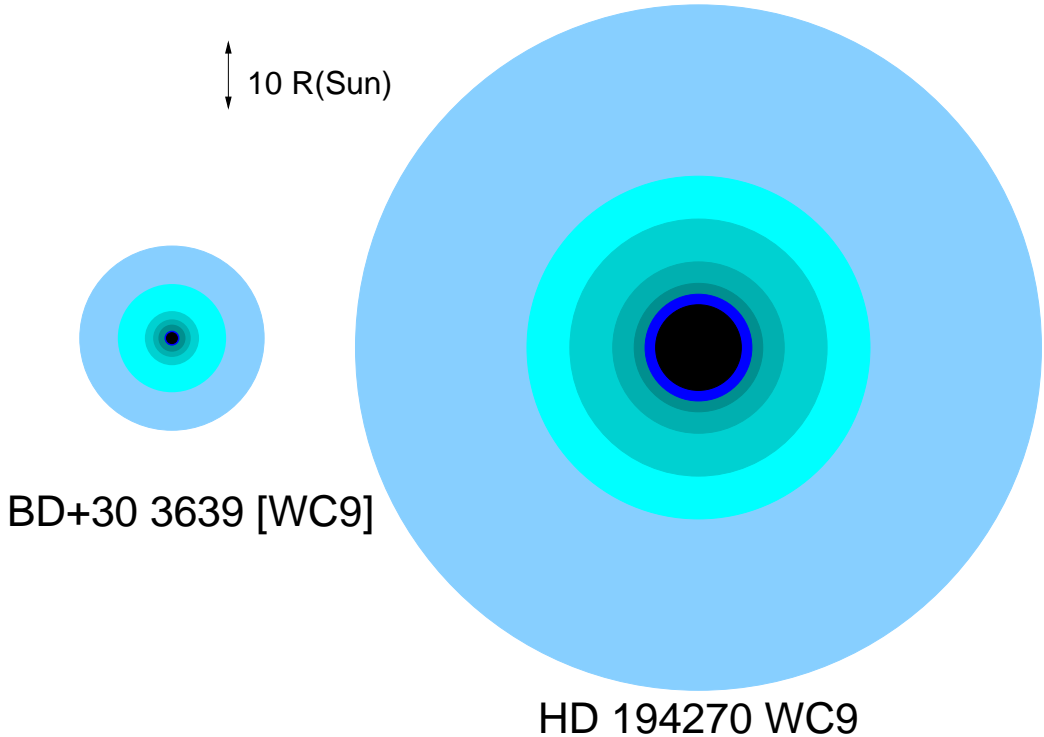}{10cm}{0}{50}{50}{-200}{0} 
\caption{Quantitative comparison between physical radii of the massive 
WC9 star HD 164270 and the [WC9]-type CSPNe BD+30$^{\circ}$ 3639.
Circles, from inner to outer, relate to $R_{\ast}$, $R_{\rm 2/3}$, 
followed by the stellar wind at
$n_{e}=10^{13}, 10^{12.5}, 10^{12}, 10^{11.5}, 10^{11}$ cm$^{-3}$. Parameters taken 
from Crowther et al. (2006),  
confirming the qualitative result of Smith 
\& Aller (1971).}
\label{wc_cspn} 
\end{figure}

Stellar temperatures for [WC] stars are difficult to characterize, 
because their geometric extension is comparable with their stellar radii. 
Atmospheric models are typically parameterized by the radius 
of the inner boundary $R_{\ast}$ at high Rosseland optical depth 
$\tau_{\rm Ross} (\sim 10$). However, only the optically thin part of the 
atmosphere is seen by the observer. The measurement of $R_{\ast}$ depends 
upon the assumption that the same velocity law holds for the visible 
(optically thin) and the invisible (optically thick) part of the 
atmosphere. 

The optical continuum radiation originates from a 
`photosphere' where $\tau_{\rm Ross} \sim 2/3$. Typical [WC] winds 
have reached a significant fraction of their terminal velocity before they 
become optically thin in the continuum. $R_{2/3}$, the radius at 
$\tau_{\rm Ross} = 2/3$ lies at highly supersonic velocities, well beyond 
the hydrostatic domain. In some weak-lined, [WO] stars, this 
is not  strictly true since their spherical extinction is modest, in which 
case R$_{\ast} \sim R_{2/3}$. 

The incorporation 
of line blanketing necessitates one of several approximations, for 
which a `super-level' approach is widely followed. 
Derived stellar temperatures depend sensitively upon the detailed 
inclusion of line-blanketing by iron-peak elements.  In general,  care 
must be taken when  comparing [WC] stars to other CSPNe on the usual 
($T_{\rm eff}, \log g$) diagrams (e.g. Werner \& Herwig 2006). 
Inferred bolometric 
corrections and stellar luminosities also depend upon detailed metal 
line-blanketing (e.g. Hillier \& Miller 1999).

Until recently, 
the number of [WC] stars studied with non-LTE, clumped, metal 
line-blanketed models has been embarrassingly small, due to the need for 
detailed,  tailored analysis of individual stars using a large number of 
free  parameters. Exceptions include Todt, Gr\"{a}fener, \& Hamann (2006) 
for [WO] stars and Crowther et al. (2003), Marcolino et al. (2007) for 
[WC] stars.
Crowther, Morris \& Smith (2006) presented a quantitative comparison 
between BD+30$^{\circ}$  3639 ([WC9]) and HD~164270 (WC9) in effect 
confirming the 
qualitative result of Smith \& Aller (1971), as illustrated in 
Fig.~\ref{wc_cspn}.

The majority of studies presented to date have focused upon optical
spectral diagnostics, often supplemented by UV datasets. Herald \& Bianchi 
(2004ab) have undertaken solely UV studies of Milky Way and LMC CSPNe 
including NGC 2371 ([WO]-PG1159) and SMP LMC 61 ([WC4]). In general
we shall concentrate upon results from optical studies, since there
may be potential systematic offsets from UV versus optical results
(as is true for massive O stars).

\section{Wind properties}

According to results from Leuenhagen et al. (1996), Leuenhagen \&
Hamann (1998) plus Koesterke \& Hamann (1997ab), using a single
stellar atmospheric code, 
late-type [WC8-10] stars possess uniformly lower wind velocities
and stellar temperatures than early-type [WC4-5] and [WO] stars, as is the 
case for massive WC stars. Typically $v_{\infty}$ = 200 -- 1000 
km\,s$^{-1}$ for late-type [WC] stars, with $T_{\ast}$ = 30 -- 80 kK
versus $v_{\infty}$ = 1200 -- 3500 km\,s$^{-1}$ and $T_{\ast}$ = 125 -- 
150 kK for early-type [WC] and [WO] stars. 

Regarding mass-loss rates,
late-type [WC] stars span a wide range i.e. $\dot{M} \sim 10^{-6 \pm 0.5}
M_{\odot}$\,yr$^{-1}$, neglecting wind clumping, while early-type
[WC] stars possess stronger winds than [WO] stars, with 10$^{-5.8 \pm 
0.3}$ and $10^{-6.3 \pm 0.3} M_{\odot}$ yr$^{-1}$, respectively.
We will show that this difference is central to early-type stars
being of either one flavour or another. Alternatively winds may
be parameterized by a so-called transformed radius, $R_{t} = R_{\ast}
(v_{\infty}/\dot{M})^{4/3}$.

It is now
well established that massive WR stars are structured, as  evidenced
from the weakness of electron scattering wings with respect to 
recombination lines (Hillier 1991) and time variable profiles (L\'{e}pine
et al. 2000; Grosdidier, Acker, \& Moffat 2000). For a volume filling 
factor $f 
(\leq$ 1), actual mass-loss rates are a factor of $1/\sqrt{f}$ lower than 
homogeneous models, although
the current approach is highly simplified. Typically $f \sim 0.1$ for both
WC and [WC] stars, so actual mass-loss rates are a 
factor of $\sim$3 lower than homogeneous results.

\section{Elemental abundances}

Late-type [WC] stars possess similar 
carbon-to-helium mass fractions
to PG1159 stars, with typically X(C)$\sim$50\% and X(He)$\sim$40\%,
in favour of the [WCL] $\rightarrow$
[WCE] $\rightarrow$ PG1159 sequence (Leuenhagen et al. 1996; Leuenhagen 
\& Hamann 1998). However, most studies presented to date for early-type 
[WC]  stars appear to show systematically {\it lower}
 carbon mass fractions of X(C)$\sim$30\% and X(He)$\sim$60\%
(Koesterke \& Hamann 1997ab; Todt et al. 2006). However, common
abundance diagnostics are not feasible for very late-type [WC] stars
(dominated by C\,{\sc ii-iii} emission lines) and early-type [WC] stars
(dominated by C\,{\sc iv} and O\,{\sc v-vi} lines). Indeed, based upon a 
narrower range of [WC] subtypes, based upon common diagnostics, Crowther  
et al. (2003)  came to quite different conclusions, preferring no
systematic differences in abundances between early-type and late-type
[WC] stars.  Marcolino et al. (2007) also emphasised the difficulties
in tightly constraining elemental abundances in [WC] stars.

\begin{figure}[ht!]
\plotfiddle{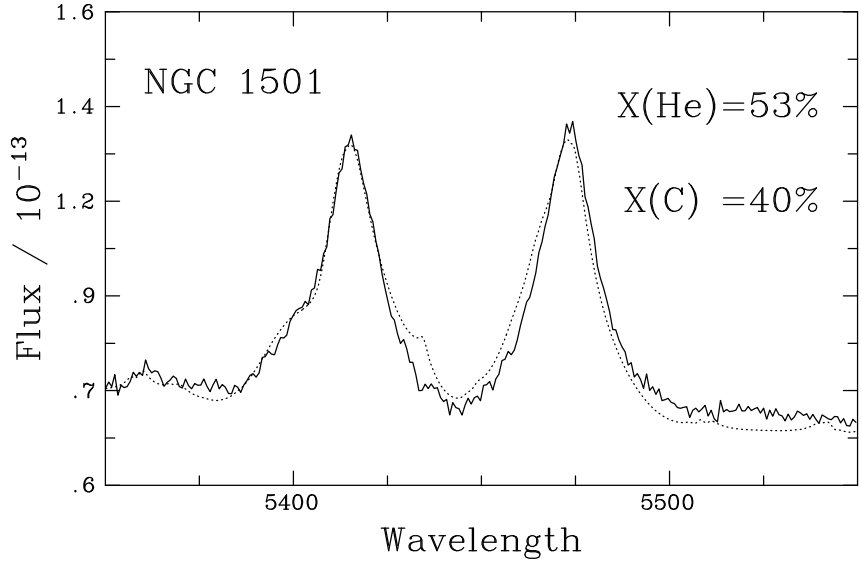}{6.5cm}{-90}{50}{50}{-200}{+230} 
\caption{Representative fit (dotted lines) to He\,{\sc ii} 5411 and 
C\,{\sc iv} 5471 lines (solid, WHT/ISIS) in NGC~1501, from which
elemental abundances X(He):X(C):X(O) = 53:40:7\% are 
obtained.}\label{ngc1501_plot} 
\end{figure}

Numerous oxygen diagnostics are available for [WC] and [WO]-type CSPNe, 
although these are highly sensitive to the precise oxygen ionization 
balance, so one should treat published abundances, typically
X(O)$\sim$10\%, as less reliable than helium or carbon. In [WC] stars, 
hydrogen has not been firmly established, although it can not be excluded 
in many CSPNe due to severe nebular contamination. There have been some
claims for a non-zero hydrogen content, such as 
X(H)$\leq$7\% by mass in He~2--113 ([WC10]) according to Leuenhagen
et al. (1996) although this was established to be of nebular origin
by De Marco, Barlow, \& Storey (1997).

Overall H, He, C and O abundances (or limits) are consistent with a 
LTP scenario. If nitrogen is detected, a very late thermal pulse (VLTP) 
would be preferred (Werner \& Herwig 2006). Due to the overwhelming carbon 
and oxygen emission line spectrum, the presence of nitrogen is 
however difficult  to confirm. Marcolino et al. suggest X(N)$\leq$0.1\% 
for all four [WC]  stars, although Todt et al. (2006) propose 
X(N)$\sim$1.5\% for PB6 ([WO]). Finally, Herwig, Langer, \& Lugaro (2003) suggest 
neutron 
capture may be responsible for the partial conversion of Fe to other heavy 
elements. From UV spectral fits, Marcolino et al. (2007) suggest 
BD+30$^{\circ}$ 
3639 is moderately  Fe-deficient, with 1/4 X(Fe)$_{\odot}$. Abell 78 
([WO]-PG1159) and LMC SMP 61 ([WC4]) are also apparently Fe-deficient 
(Herald \& Bianchi 2004ab).

\section{New results for [WC] stars}

In the light of apparently conflicting results, and the differences
in subtype distributions of [WC]-type CSPN and massive WC stars
(Fig.~\ref{wc_pop}), 
a new study of [WO], early- and late-type [WC] CSPNe has been undertaken. 
This is based upon a sample of 9 CSPNe, limited to flux calibrated optical 
spectroscopy from INT/IDS (NGC~6905, NGC~6751), WHT/ISIS (NGC 1501, 
M1--25, M2--43, NGC 40, He~2 459, BD+30$^{\circ}$ 3639) and AAT/RGO (Sand 
3).
It was necessary to employ flux calibrated spectroscopy to account
for the underlying nebular continuum. In common with Todt et al. (2006)
we adopt $\log (L/L_{\odot}) = 3.7 $ and a mass of $0.6 M_{\odot}$.
For the spectral analysis, the CMFGEN code is employed, for which
line blanketing by He, C, O, Ne, Mg, Si, S, Ar, Ca, Fe and Ni are
included (see Crowther et al. 2006 for details). 

We have also re-examined the physical properties of CPD-56$^{\circ}$ 8032 
([WC10]) based on line blanketed models, and confirm the results of De 
Marco \& Crowther (1998), adjusted for a clumped wind, with the exception 
that use of an extended C\,{\sc ii} model atom (P.J. Storey) suggests a 
lower mass fraction of X(He):X(C):X(O) $\sim$ 50:30:20\%. Formally, we 
omit 
CPD-56$^{\circ}$ 8032 from our study given our desire to employ common 
abundance diagnostics for the entire sample.

\begin{table}[!ht]
\caption{Physical and wind properties of Galactic Wolf-Rayet 
CSPNe, adopting $\log (L/L_{\odot})$ = 3.7 (0.6 $M_{\odot}$) and
wind clumping with $f$ = 0.1.}\label{new}
\smallskip
\begin{center}
\begin{tabular}{l@{\hspace{2mm}}c@{\hspace{2mm}}c@{\hspace{0mm}}
c@{\hspace{0mm}}c@{\hspace{2mm}}c@{\hspace{3mm}}l}
\tableline
\noalign{\smallskip}
Subtype & $T_{\ast}$ & $R_{\rm 2/3}$ & $\log \dot{M}$ & $v_{\infty}$ & 
Bol.Corr. & Sample\\
        & kK & $R_{\odot}$& $M_{\odot}$ yr$^{-1}$ & km\,s$^{-1}$ & 
mag &\\
\noalign{\smallskip}
\tableline
\noalign{\smallskip}
{}[WC10]  & 33  & 2.5 & --6.0 & 225 & --2.4 &CPD--56$^{\circ}$ 8032 \\
{}[WC9] & 55--65 & 0.8$\pm$0.15 & --6.1 & 700--1600 & 
--3.4$\pm$0.4 & BD+30$^{\circ}$ 3639\\ 
        & & & & & & He~2-459\\
{}[WC8] & 80--85 & 0.45$\pm$0.05 & --6.3 & 950--1100 & --4.0 & NGC 40, 
M2--43\\
{}[WC4--5] & 110--145 & 0.3$\pm$0.15 & --6.3 & 1100--2350 & --4.7$\pm$0.5 
&M1--25, NGC 6751\\
{}[WO1--4] & 115--150 & 0.16$\pm$0.04 & --7.0 & 1900--2500 & --6.4$\pm$0.2
& NGC~1501,\\
           &          &          &   &           && NGC~6905, Sand 3\\
\noalign{\smallskip}
\tableline
\end{tabular}
\end{center}
\end{table}

In common with previous studies we confirm increased stellar temperatures
and wind velocities from [WC10] to [WC8--9] to [WC4--5], with again no 
significant differences between [WC4--5] and [WO1--4] stars, as indicated 
in
Table~\ref{new}. We conclude that the primary 
differences between [WC4--5] and [WO1--4] stars results from differences
in mass-loss rate. After allowing for wind clumping, we find
$\dot{M} \sim 5 \times 10^{-7} M_{\odot}$\,yr$^{-1}$  for [WC4--5]
stars, versus $1 \times 10^{-7} M_{\odot}$\,yr$^{-1}$ for [WO1--4] stars.
For massive WC and WO stars, earlier spectral types also follow  from 
reduced wind densities for otherwise identical physical properties.

\begin{figure}[ht!]
\plotfiddle{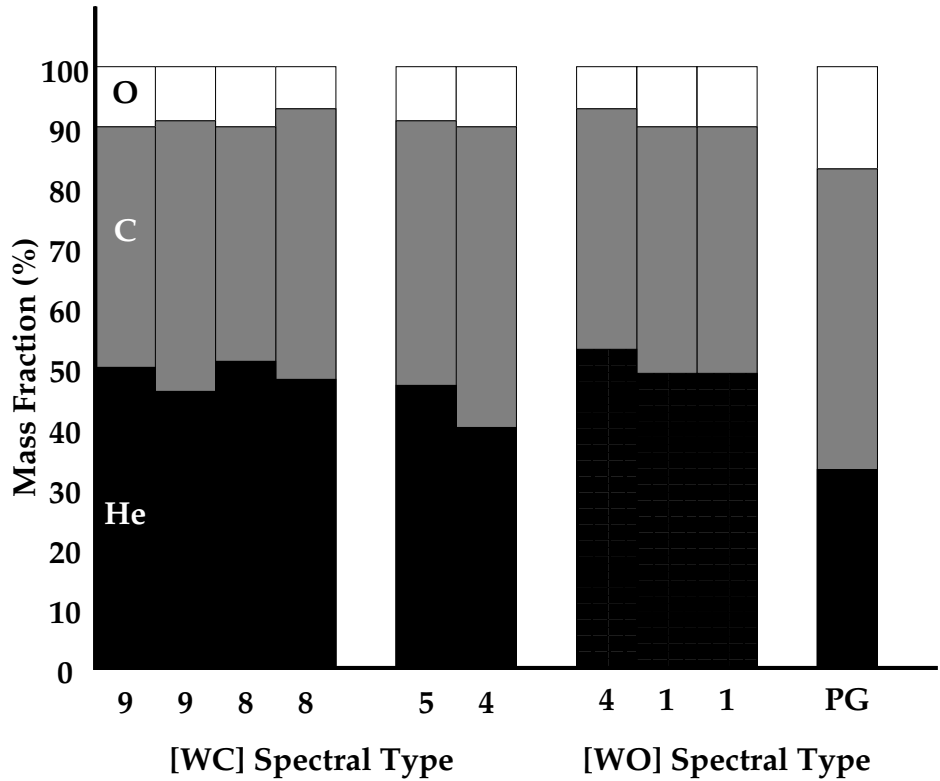}{10cm}{0}{80}{80}{-175}{0} 
\caption{Elemental abundances for late- and early-type 
[WC] and [WO] stars from the present study, together with typical
PG1159 star abundances (labelled PG). He, C and O mass fractions
are shown in black, grey and white, respectively.
}\label{crowther_final_mono}
\end{figure}

Crowther (1999) noted that {\it strong} winds cause {\it weak} O\,{\sc vi} 
emission (WC4--5 subtypes) in
hot carbon-rich WR stars as a result of recombination to lower ionization
stages of oxygen in the optical line formation region 
(10$^{11} \leq n_{e} \leq 10^{12}$ cm$^{-3}$). Conversely,
{\it weak} winds permit {\it strong} O\,{\sc vi} emission (WO subtypes) 
since oxygen does not recombine to lower ionization stages until
lower electron densities. Emission lines from [WO]-type CSPN are typically 
narrower than for early-type [WC] stars for similar reasons, in the
sense that winds have not yet reached their maximum velocities in the
optical line forming region of [WO] stars, in contrast to [WC] stars.

Wind densities of WR stars also affects the emergent ionizing flux 
distributions, in the sense that {\it soft} fluxes result from 
strong-lined stars with dense winds, with {\it hard} fluxes for weak-lined 
stars with lower density winds (Schmutz, Leitherer, \& Gruenwald 1992; 
Smith, Norris, \& Crowther 2002). Consequently, bolometric corrections for 
[WC] stars are functions of wind 
density as well as stellar temperature (Table~\ref{new}). Use of 
appropriate spherical atmospheric models is relevant to photoionization 
models of [WC] stars, such as Ercolano et al. (2004) who made use of 
plane-parallel models for NGC~1501 ([WO4]).

For NGC~40 ([WC8]) and NGC~6905 ([WO4]) identical INT/IDS 
optical  spectroscopy was used to the study of Marcolino et al. (2007).
In general, differences in physical and wind properties are modest, with 
the exception of elemental abundances. We rely solely upon fits to 
He\,{\sc ii}
5411/C\,{\sc iv} 5471 throughout (Hillier 1989), as illustrated 
in Fig.~\ref{ngc1501_plot} for NGC~1501. Marcolino et al. choose to
employ  the whole spectral region for their abundance estimates, although
this yields poor fits to these diagnostics in some instances. For 
Sand~3 ([WO1]) our results are intermediate between Todt et al. (2006) and 
recombination line studies (Barlow \& Hummer 1982), while both
model atmosphere results differ from recombination line results in
NGC~1501 ([WO4]), as illustrated in Table~\ref{Sand3}. 

\begin{table}[!ht]
\caption{Stellar elemental abundances for [WO] stars
Sand~3 and NGC~1501 (by mass in \%).}\label{Sand3}
\smallskip
\begin{center}
\begin{tabular}{cccl}
\tableline
\noalign{\smallskip}
X(He)  & X(C) & X(O) & Study \\
\noalign{\smallskip}
\tableline
\noalign{\smallskip}
\multicolumn{4}{c}{Sand 3} \\
38 & 54 & 8 & Barlow \& Hummer 1982 \\
62 & 26 & 12 & Todt et al. 2006 \\
49 & 41 & 10 & This study \\
\noalign{\smallskip}
\multicolumn{4}{c}{NGC 1501} \\
\noalign{\smallskip}
36 & 48 & 16 & Ercolano et al. 2004 \\
55 & 35 & 10 & Todt et al. 2006 \\
53 & 40 & 7 & This work \\
\noalign{\smallskip}
\tableline
\end{tabular}
\end{center}
\end{table}

In common with Crowther et al. (2003) and Marcolino et al. (2007) we 
confirm no {\it systematic } difference between abundances in late-type 
and early-type [WC] and [WO] stars, as illustrated in 
Fig.~\ref{crowther_final_mono}. Regardless of the differences between our 
results and other studies, observations of the peak intensities of the 
He\,{\sc ii} 5411 and C\,{\sc iv} 5471 across our sample favour a narrow 
range in elemental abundances.

\section{Evolutionary connections with other H-deficient stars}

\begin{figure}[ht!]
\plotfiddle{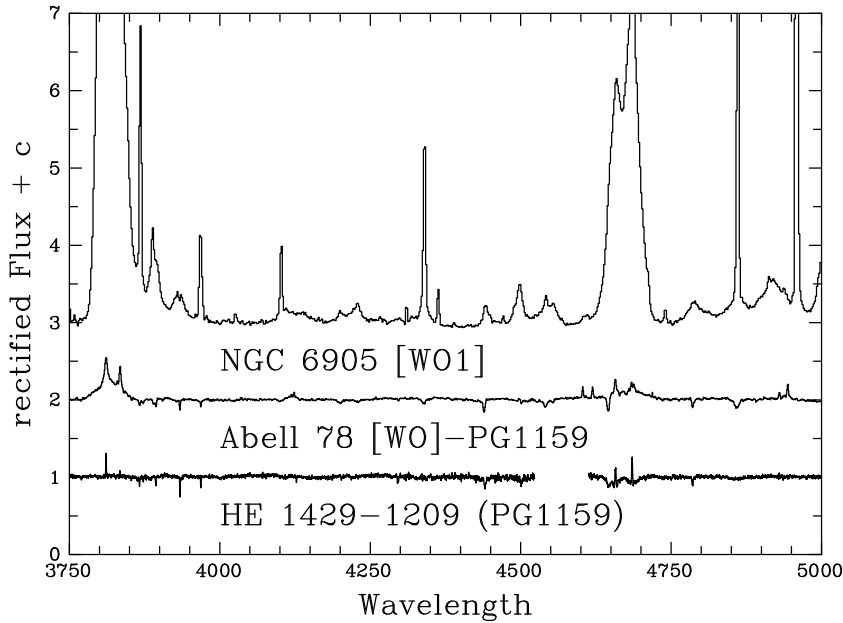}{7.5cm}{-90}{50}{50}{-200}{+250} 
\caption{Spectral comparison between HE 1429--1209 (PG1159),
Abell 78 ([WO]-PG1159) and NGC 6905 [WO1]). Derived
elemental abundances are X(He):X(C):X(O) = 38:54:6, 54:37:8, 49:41:10\%
respectively, according to Werner et al. (2004), Herald \& Bianchi
(2004a) and this work.}\label{wc_pg1159} 
\end{figure}

If we adopt $M = 0.6 M_{\odot}$ for early-type [WC] and [WO]
CSPNe together  with 
the stellar radii inferred from $T_{\ast}$ we obtain $\log g \sim $ 6, 
versus $\log g \sim$7 for PG1159 stars. Overall, common abundances
(recall Fig.~\ref{crowther_final_mono}),
plus a continuum of decreasing wind strengths and increasing 
surface  gravities from early-type [WC], [WO], [WO]-PG1159, and PG1159 
suggests a direct evolutionary sequence, as illustrated in 
Fig.~\ref{wc_pg1159}. Indeed, Longmore~4 (PG1159)
has undergone a number of brief outbursts in which a modest
increase in wind density caused a [WO]-PG1159 appearance (Bond,
these proc.).

With regard to [WC8--9] stars, again adopting
$M = 0.6 M_{\odot}$ together  with 
the stellar radii inferred from $T_{\ast}$ we obtain $\log g \sim $ 
4.5--5. Indeed, common abundances with [WC4--5] and [WO]
CSPNe suggests an common evolutionary sequence, apart for the
well-known gap between [WC8] and [WC5] in the subtype distribution
for Galactic disk [WC]-type CSPNe (Fig.~\ref{wc_pop}).
Physically, little distinguishes the
properties of such stars, with the exception of a modest
decrease in wind density from [WC8] to [WC5], according to Crowther
et al. (2002). This difference is almost certainly attributable to the 
higher surface gravities of CSPN with respect to their massive 
counterparts, resulting in substantially earlier subtypes on average.
If we adopt the usual mass-luminosity relation for Galactic
WC5--9 stars, we find $\log g \sim$4.5--5 in all cases, such that
spectral types are primarily due to subtle differences in wind density
and stellar temperature. Indeed, weaker winds in massive LMC WC stars
results in universal WC4 or WO spectral types (Crowther et al. 2002).

\section{Summary}

We discuss physical and wind properties of Wolf-Rayet CSPNe drawn from
the recent literature plus new analyses for a range of subtypes. 
In general the higher ionization lines seen at earlier spectral type
indicates an increased stellar temperature, although very early [WO]
subtypes  are favoured for hot CSPNe with weak winds, with [WC4--5]
subtypes resulting for hot CSPNe with stronger winds. [WC8--9] subtypes
correspond to lower temperature CSPNe with strong winds, with much
lower temperatures indicated in [WC10] stars. We explain the differences
in subtype distribution between Galactic disk CSPNe and massive WC stars
as a result of decreased wind densities in the former, owing to increased
surface gravities. Massive WC stars  in the LMC differ from  Galactic WC 
stars through reduced wind densities caused by lower metallicities, 
rather than increased surface gravities.

There does {\it not} appear to be 
a systematic difference between carbon-to-helium mass fractions in
late to early-type CSPNe, at least for subtypes for which we are able to
employ common diagnostics ([WC9] to [WO1]). Consequently, evolution
from late-type [WC] through early-type [WC] and [WO] to PG1159 stars
appears to be consistent with most abundance patterns, and expectations
for a late thermal pulse. Indeed, we note that similar analysis tools 
have been applied to the post He-flash system V605 Aql, which has now 
evolved through to a early-type [WC] spectral type over the past 80 years
and shares a similar abundance pattern with X(He):X(C):X(O) = 54:40:5\%
(Clayton et al. 2006).

%\subsection{}   %%% Second level section head (remove "%" symbol)
%\subsubsection{}   %%% Lowest level section head (remove "%" symbol)
%\section*{}    %%% Unnumbered top level section head (remove "%" symbol)
%\subsection*{}   %%% Unnumbered second level section head (remove "%" symbol)

\acknowledgements %%% Text of acknowledgements runs on after this command.
Financial support from the workshop organisers was greatly appreciated. 
Thanks to Orsola De Marco and Mike Barlow, from whom INT and AAT 
spectroscopy of [WC]
and [WO] stars were obtained. Klaus Werner kindly provided the spectrum of 
HE~1429--1209.

%%% THE BIBLIOGRAPHY
%%%
%%% CONSULT SECTION 3 OF "INSTRUCTIONS FOR AUTHORS" FOR HOW TO USE NATBIB.
%%% AUTHORS ARE ENCOURAGED TO USE EITHER THE "THEBIBLIOGRAPY" ENVIRONMENT
%%% BY UNCOMMENTING (DELETING THE "%" SYMBOL) THE COMMANDS BELOW, OR BY
%%% USING THE BIBTEX ENVIRONMENT. TO FIND OUT WHICH IS APPLICABLE TO YOUR
%%% CONTRIBUTION, CONSULT THE VOLUME EDITORS FOR YOUR PROCEEDINGS.
%%%

\end{document}